\documentclass[multphys,vecphys]{svmult}

\usepackage{makeidx}     % allows index generation
\usepackage{graphicx}    % standard LaTeX graphics tool
\usepackage{multicol}    % used for the two-column index
\usepackage{amsmath}
\makeindex             % used for the subject index
\begin{document}

\title*{A most energetic type Ic supernova: SN 2003L}
% Use \titlerunning{Short Title} for an abbreviated version of
% your contribution title if the original one is too long
\author{Alicia M. Soderberg\inst{1}}
% Use \authorrunning{Short Title} for an abbreviated version of
% your contribution title if the original one is too long
\institute{Palomar Observatory, 105-24, California Institute of Technology, Pasadena, CA 91125
\texttt{ams@astro.caltech.edu}}
%
% Use the package "url.sty" to avoid
% problems with special characters
% used in your e-mail or web address
%
\maketitle

\begin{abstract}
We present extensive radio observations of SN 2003L, the most luminous
and energetic Type Ic radio supernova with the exception of SN 1998bw.
Using radio data, we are able to constrain the physical
parameters of the supernova, including the velocity and energy of the
fastest ejecta, the temporal evolution of the magnetic field, and the
density profile of the surrounding medium.  
We highlight the extraordinary properties of the
radio emission with respect to the supernova's normal characteristics
within optical bands.  We find that although the explosion does
not show evidence for a significant amount of relativistic ejecta, it
produces a radio luminosity which is comparable to that seen in the
unusual SN 1998bw.  Using SN 2003L as an
example, we comment briefly on the broad diversity of type Ic
properties and the associated implications for progenitor models.
\end{abstract}

\section{Introduction}
\label{sec:1}
Despite active campaigns to study radio emission from type Ib/c
supernovae, only a small number of events have been successfully
detected.  Among the class of radio bright supernovae is SN 1998bw, an
unusually bright type Ic discovered within the error box of the nearby
gamma-ray burst GRB 980425. Reaching a peak radio luminosity $\sim100$
times higher than all other radio bright type Ib/c supernovae (SNe),
it has been proposed that SN 1998bw was powered by a central engine, 
similar to the popular model for gamma-ray bursts (GRBs) [3].  

In this paper we present observations of the first radio bright type
Ib/c supernova with energetics comparable to those shown in SN 1998bw.
SN 2003L was optically discovered on 2003 Jan 12.15 UT [2] and spectroscopically identified on 2003 Jan 25.0 UT
[5,8]. The supernova was
seen to bear strong resemblance to the typical type Ic SN 1994I at
maximum light, showing low average expansion velocities of 5900 km/s as derived
from the Si II line.  The optical light-curve peaks at $m_V\approx 16$ which
places SN 2003L among the brightest optical type Ic SNe observed to date.

%\begin{equation}
%\vec{a}\times\vec{b}=\vec{c}
%\end{equation}

\section{Observations with the VLA}
On Jan 26.23 2003 UT we detected a radio transient coincident with the
optical position of SN 2003L.  We subsequently initiated an intense
radio monitoring campaign at the Very Large Array (VLA) to trace the
evolution of the radio emission from the supernova.  Data were taken
in standard continuum observing mode with a bandwidth of $2\times 50$
MHz centered on frequencies 8.5, 15.0 and 22.5 GHz.  Flux
density measurements were derived using calibrator 3C286 and phase
referenced against calibrators J1118+125, J1120+134, and J1103+119.
Data were reduced using standard packages within the Astronomical
Image Processing System (AIPS).  
At 8.5 GHz (our most densely sampled light-curve)
typical flux uncertainties were $\sim 60~\mu$Jy for an average
integration time of 10 minutes. 
The results of our radio monitoring campaign of SN 2003L are summarized as
Figure 1. These observations demonstrate a broad spectrum, similar to that
observed for SN 1998bw.
\begin{figure}
\centering
% Use the relevant command for your figure-insertion program
% to insert the figure file.
% For example, with the option graphics use
\includegraphics[height=5cm]{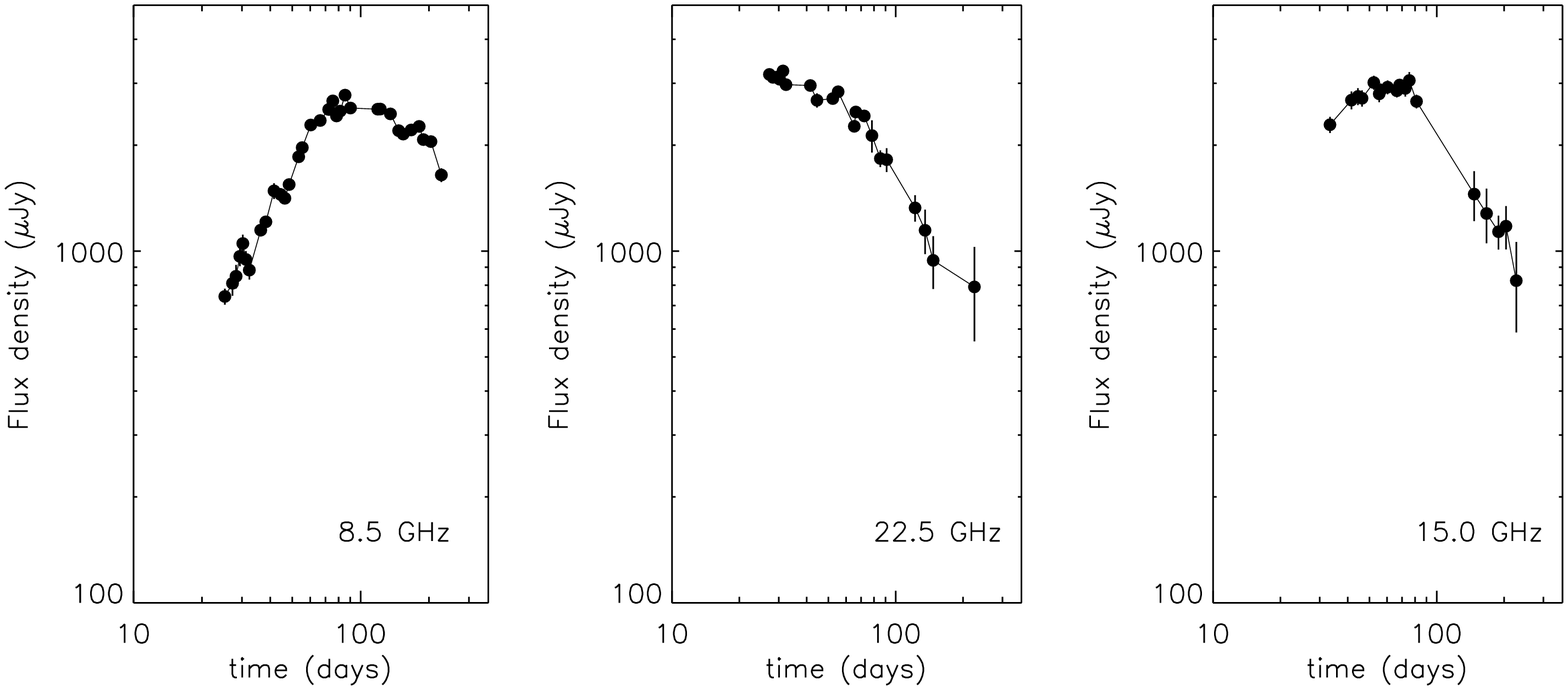}
%
% If not, use
%\picplace{5cm}{2cm} % Give the correct figure height and width in cm
%
\caption{Radio light-curves for type Ic SN 2003L at 8.5, 15.,
and 22.5 GHz for the time period 2003 Jan 26 - 2003 Aug 15 UT.}  
\label{fig:1}       % Give a unique label
\end{figure}
\section{Robust Constraints}

Here we discuss the constraints imposed by the  
radio observations for SN 2003L.
As a preliminary constraint on the total energy of the source, we
estimate the brightness temperature ($T_b$) of the supernova and compare it
with the robust constraints imposed by equipartition arguments and the
inverse Compton catastrophe (ICC). From [6], the brightness
temperature is defined as a function of the 
observed flux density, the peak frequency, and the angular size
of the source.
As an initial estimate for the physical size of the SN ejecta, we first
assume the optical expansion velocity of $5900\rm ~km~s^{-1}$
[8] can be used as an average speed to describe the
motion of the radio bright ejecta.  Using an approximate explosion
date of Jan 1 2003 UT based on optical light-curves [7], 
we estimate the shock radius to be $r\approx4.3\times
10^{15}$ cm at $t\approx85$ days when the observed flux density peaked
at 8.5 GHz.  Using the observed peak flux of $F_{8.5\rm
~GHz}\approx 2.8$ mJy and adopting a source distance of 91.7 Mpc
($\Omega_M=0.27$, $\Omega_{\Lambda}=0.73$, $H_0=71~\rm
km~s^{-1}~Mpc^{-1}$), we find a brightness temperature of $T_b\approx
1.8\times10^{12}$ K which is dangerously near the 
ICC limit of $T_b < 10^{12}$ K.  This suggests that the
radio ejecta expanded with a velocity significantly higher than that
observed at optical frequencies.  In fact, an ejecta velocity of $\sim
16,000$ km/s ($\sim 3$ times larger than that derived from optical
spectroscopy) would be necessary to avoid violating the ICC
constraint.

Assuming equipartition of energy places a further constraint on the
brightness temperature limit and reduces it to $T_b < 5\times
10^{10}$ K.  Using the equipartition arguments of [6] and
[3], we derive the minimum energy for the radio supernova.
Assuming that the observed radio flux is produced by synchrotron
emission, the total energy of the source ($U$) can be expressed as the
sum of the energy in relativistic electrons ($U_{\rm e}$) and the
energy in the magnetic field ($U_{\rm B}$).  At equipartition, the
fraction of total energy in electrons equals the fraction of total
energy in magnetic fields ($\epsilon_e=\epsilon_B=1$) and the total
energy is minimized at $U_{\rm eq}$ [6]. This occurs when
the emitting source reaches an equipartition radius denoted by the
angular size, $\theta_{\rm eq}$.  The minimum energy of the source can
be thus be parameterized in terms of the synchrotron peak flux and the equipartition size.

Using our most densely sampled light-curve, we fit for the peak flux ($S_p$)
over the observed $\sim 200$ day evolution and find $S_p\approx 2.8$
mJy at $\nu_p=8.5$ GHz on Mar 27 2003 UT ($\approx 85$ days since
explosion).  For this epoch we estimate an angular size $\theta_{\rm eq}\approx 19 \mu$as (with
$\beta\approx -1.0$) 
which implies an average shock velocity of $v\approx 0.1c$ and an
equipartition brightness temperature of $T_{b_{\rm eq}}\approx
5.0\times 10^{10}$ K. By setting $U=U_{\rm eq}$ we find
the energy is minimized at the equipartition value of $U_{\rm
eq}\approx 4.3\times 10^{47}$ erg with an associated magnetic field 
strength of $B_{p_{\rm eq}}\approx 0.6\rm ~G$.
As shown by [6], synchrotron emission systems which diverge from equipartition
necessitate a huge increase in total energy. Consequently, it is
possible that the total energy contained within the fast moving ejecta
of SN 2003L is in fact much larger than $4.3\times 10^{47}$ erg.

These preliminary constraints allow us to make two robust conclusions:
1.) the velocity of the radio bright SN ejecta must be at least
$16,000~\rm km~s^{-1}$ to avoid violating the inverse Compton
catastrophe limit and $\sim 30,000~\rm  km~s^{-1}$ assuming equipartition,
2.) the energy of the supernova must be $> 4.3\times 10^{47}$ erg and
could be significantly larger depending on the proximity of the system
to equipartition.  These conclusions imply that there was a
considerable amount of energy released at high velocities in the type
Ic supernova explosion of SN 2003L.  In fact, these equipartition
constraints alone demand that SN 2003L is among the most energetic type
Ib/c supernovae observed to date, second only to the unusual event of
SN 1998bw/GRB 980425.  Figure 2 is a compilation of all the radio bright
type Ib/c supernovae observed to date.  By comparing the peak radio
luminosity to the observed time of peak flux, the diversity in
equipartition derived expansion velocities can be examined.  
Note that although SN 2003L peaks
later, it is among the brightest radio supernovae.
\begin{figure}
\centering
% Use the relevant command for your figure-insertion program
% to insert the figure file.
% For example, with the option graphics use
\includegraphics[height=7cm]{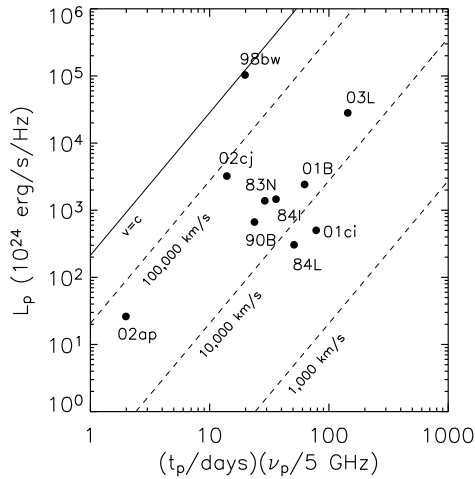}
%
% If not, use
%\picplace{5cm}{2cm} % Give the correct figure height and width in cm
%
\caption{The diversity of peak luminosities and observed time of the
peak is shown above for all radio bright type Ib/c supernova.  By
assuming equipartition, the expansion velocity for each event can be
estimated (dashed lines).Note the radio ejecta of SN 2003L is relatively slow
and luminous.}
\label{fig:1}       % Give a unique label
\end{figure}
\section{Implications and Physical Parameters}
Assuming equipartition values of $\epsilon_B=\epsilon_e=1$ and
adopting synchrotron self-absorption as the dominant absorption process,
we determine the temporal evolution of the total energy and radius of 
the ejecta for SN 2003L.
For observations spanning $t=25-200$ days, the total energy increases from
$E\approx3.0\times 10^{47}$ to $1.1\times 10^{48}$ erg. Over the same
period, the shock radius scales as $r\propto t^{0.67}$ from $r\approx
1.45\times 10^{16}$ to $5.0\times 10^{16}$ cm.  Assuming an explosion
date of Jan 1 2003 UT, the average velocity decreases from
$v\approx0.22c$ to $0.10c$. 
In comparison to SN 1998bw where it was observed that $v\approx c$, it
is clear that SN 2003L does not have a significant amount of material
moving at relativistic speeds.  However, the total energy within the radio
bright ejecta remains quite high at $\sim 10\%$ that of SN 1998bw (on similar
time-scales) and 10-100 times greater than other radio supernovae.
 
Using the values derived for the total energy and radius, we predict the
magnetic field decreased from $B\approx 0.75$ G  to $0.23$ G over the period
$t=25-200$ days.  We find an temporal evolution of $B\propto
t^{-0.63}$, such that $B\propto r^{-0.93}$.  Extrapolating, we find 
$B(t)\approx 10$ G at$r\approx 10^{15}$ cm.  For comparison, type IIb 
SN1993J exhibited a
similar evolution with $B\approx 60$ G at $~r\approx 10^{15}$ cm
and a radial scaling of $B(r)\propto r^{-1}$.
 
Environmental properties can also be predicted based on our radio
observations.  From 25 days to 200 days, we find that the electron
density, $n_e(t)$, drops by a factor of $\sim 10$ from $n_e\approx
340\rm ~cm^{-3}$ to $36\rm ~cm^{-3}$.  Expressed in terms of a radial
dependence, $n_e\propto r^{-1.8}$; similar to the density profile
expected from a massive stellar wind, $n_e\propto r^{-2}$.  This was also
seen in the case of SN 1998bw/GRB980425 [4], although solid evidence
for a wind environment has yet to be detected for the majority of
observed gamma-ray bursts.
 
Directly coupled to the electron number density is the mass loss rate,
$\dot{M}$ of the progenitor star.  For an assumed stellar wind velocity of 
$w=10^3~\rm km~s^{-1}$, we find a roughly constant mass loss rate of
$\dot{M}(t)\approx 5\times 10^{-6}~\rm M_{\odot}~yr^{-1}$.  This is
$\sim$10 percent larger than values derived for SN 1998bw and SN 2002ap,
of $\dot{M}(t)\approx 2.5\times 10^{-7}$ and $5\times 10^{-7}~\rm
M_{\odot}~yr^{-1}$, respectively [1,4].  It should be noted that these rates 
are consistently smaller than the predicted mass loss rate
for Wolf-Rayet stars ($\dot{M}(t)\approx 10^{-4}-10^{-5}~\rm
M_{\odot}~yr^{-1}$) , which are thought to be the progenitors of type
Ic supernovae.

\section{Discussion}

Although the constraints provided by equipartition arguments are
robust, they are also preliminary.  The extensive radio data set for
SN 2003L warrants a full modeling effort to accommodate the effects of
multiple absorption processes including synchrotron self-absorption
and free-free absorption.  Results from our radio modeling study of
SN 2003L will be presented along with extensive broadband data for this
event [7].  We will show that the highly energetic supernovae SN 2003L is 
becoming one of the best studied radio supernovae to date, thereby offering new
insights on the diversity of cosmic explosions.

% BibTeX users please use
% \bibliographystyle{}
% \bibliography{}
%
% Non-BibTeX users please follow the syntax
% the syntax of "referenc.tex" for your own citations
%%%%%%%%%%%%%%%%%%%%%%%% referenc.tex %%%%%%%%%%%%%%%%%%%%%%%%%%%%%%
% sample references
% "physics"
%
% Use this file as a template for your own input.
%
%%%%%%%%%%%%%%%%%%%%%%%% Springer-Verlag %%%%%%%%%%%%%%%%%%%%%%%%%%

%
% BibTeX users please use
% \bibliographystyle{}
% \bibliography{}
%
% Non-BibTeX users please use

%%%%%%%%%%%%%%%%%%%%%%%%%%%%%%%%%%%%%%%%%%%%%%%%%%%%%%%%%%%%%%%%%%%%%%  }

%%%%%%%%%%%%%%%%%%%%%%%%%%%%%%%%%%%%%%%%%%%%%%%%%%%%%%%%%%%%%%%%%%%%%%

\printindex
\end{document}